\documentclass[a4paper,11pt]{article}
\usepackage{color}
\usepackage{amsmath}    
\usepackage{amssymb}
\usepackage{graphicx}   
\usepackage{subfigure}
\usepackage{longtable}
\usepackage[utf8]{inputenc} 
\usepackage[margin=3.2cm]{geometry}
\title{Inherent directionality explains the lack of feedback loops in
  empirical networks} 

\author{Virginia Dom{\'\i}nguez-Garc{\'\i}a , $^{1}$, Simone Pigolotti, $^{2}$, Miguel A. Mu\~noz$^{1,\ast}$\\
  \normalsize{$^{1}$ Departamento de Electromagnetismo y F\'isica de la Materia,}\\
  \normalsize{and Instituto Carlos I de F\'isica Te\'orica y Computacional}\\
  \normalsize{ Universidad de Granada, 18071 Granada, Spain.}\\
  \normalsize{$^{2}$  Departament de Fisica i Enginyeria Nuclear, Universitat Politecnica de Catalunya}\\
  \normalsize{Rambla Sant Nebridi s/n, 08222 Terrassa, Barcelona, Spain.} \\
  \small{$^{*}$Corresponding author: mamunoz@onsager.ugr.es}}

\begin{document}     

\maketitle
\begin{abstract} 
  We explore the hypothesis that the relative abundance of feedback
  loops in many empirical complex networks is severely reduced owing
  to the presence of an inherent global directionality.  Aimed at
  quantifying this idea, we propose a simple probabilistic model in
  which a free parameter $\gamma$ controls the degree of inherent
  directionality.  Upon strengthening such directionality, the model
  predicts a drastic reduction in the fraction of loops which are also
  feedback loops.  To test this prediction, we extensively enumerated
  loops and feedback loops in many empirical biological, ecological
  and socio- technological directed networks. We show that, in almost
  all cases, empirical networks have a much smaller fraction of feedback
  loops than network randomizations. Quite remarkably, this empirical
  finding is quantitatively reproduced, for all loop lengths, by our
  model by fitting its only parameter $\gamma$. Moreover, the fitted
  value of $\gamma$ correlates quite well with another direct
  measurement of network directionality, performed by means of a novel
  algorithm.  We conclude that the existence of an inherent network
  directionality provides a parsimonious quantitative explanation for
  the observed lack of feedback loops in empirical networks.
\end{abstract}

\section*{Introduction}

Genetic regulatory circuits, metabolic pathways, food webs, and many
different socio-technological systems can be visualized as networks
made up of units linked pairwise whenever there is some sort of
``interaction'' or ``flow'' between them.  In many cases, empirical
networks are dynamical, time-changing entities, and most of the
existing compiled datasets represent static snapshots or time-averages
over some observation interval of these more complex
processes. Nevertheless, the description in terms of static networks
has proven to be useful to identify structural features which are
responsible for emerging functions
\cite{Newman-review,Barabasi,Vespi,Arenas_syncro}.  Some structural
features, including clustering, degree assortativity
\cite{assortativity}, and the relative abundance of specific motifs
\cite{Alon-motifs,Alon-book}, characterize the topology at the local
scale. Other traits, such as nestedness
\cite{Nestedness,plos_nestedness}, community structure
\cite{Newman_compartments,Stouffer_compart}, and the existence of a
hierarchy \cite{hierarchy_barabasi,Sole} are related to the
large-scale organization.  Clearly, these features are not necessarily
independent.

In many empirical networks, interactions are directed, i.e. links have an
origin and a target node.  This direction can be generally thought of
in terms of flows, such as the energy transfer in food webs
\cite{Dunne} and the flow of biological information in genetic or
neural networks.  Often, this flow identifies a global inherent
directionality. By ``inherent directionality'' we mean that all nodes
can be ordered on a one-dimensional axis, in such a way that links
point preferentially from low to high values of their coordinates in
such an axis. In this sense, the existence of an inherent
directionality is deeply related to the existence of a hierarchical
structure \cite{Ulanowicz,Sole}.  For example, ({\it i}) in networks
where there is a transfer of matter, such as food webs or metabolic
networks, one can identify a hierarchy of ``trophic'' levels (links
tend to point from lower levels to higher ones), ({\it ii}) in gene
regulatory networks there is a hierarchy of control (controller nodes
act upon controlled ones), and ({\it iii}) in neural networks, the
flow of information propagates from sensory neurons at the bottom of
the hierarchy, to neurons in the central system at intermediate
levels, and from there to the level of motor neurons.

The existence of an inherent directionality can have a deep impact on
the network small-scale structure, in particular on the statistics of
motifs, such as feedback loops.  In a directed network, a ``feedback
loop'' of length $k$ is defined as a closed sequence of $k$ different
nodes in which a walker following the directions of the arrows returns
to the starting point after visiting once and only once all $k$
nodes. Feedback loops are well-known to have a profound impact on
dynamical stability in food webs
\cite{Neutel2002,Neutel2007,McCann,allesina_2008,May2009,complexity,100_questions,
  mitchell_neutel_2012} as well as in biological and generic networks
\cite{dambacher,levins_74,math1,math2,Alon-book,Mangan,Angeli,Thomas,Plathe,Tyson2008,Simone_loops,simone_oscilations,Serrano}.
``Structural loops'' or simply ``loops'', defined as closed sequences of
pairwise connected nodes, independently of the direction of links are
also of interest. Clearly, the set of feedback loops is a subset of
that of structural loops.

The relationship between the existence of a inherent directionality
and feedback loops can be intuitively understood by considering the
case of perfect directionality --or feedforwardness-- in which all
links are aligned with the inherent directionality. In such perfectly
directional networks, feedback loops are completely absent, as at
least one link against the directionality is required to close a
feedback loop.  The impact of directionality on the statistics of
feedback loops is less trivial to assess in cases of incomplete
feedforwardness, where directionality only partially determines the
direction of links.

\begin{figure}[ht!bp]
\begin{center}
\includegraphics[width=0.7\textwidth]{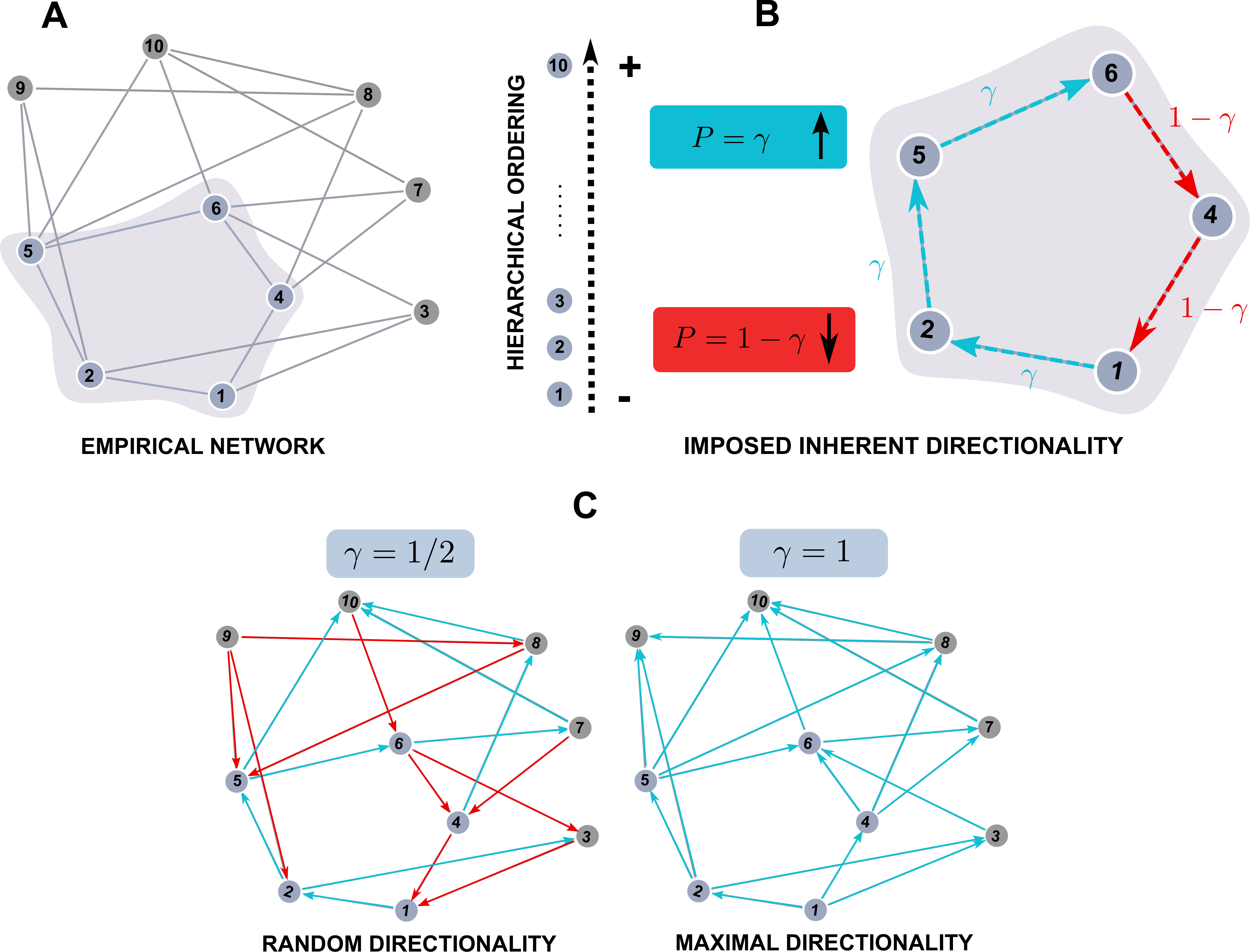}
\caption{ 
{\bf Schematic representation of the directionality
    model}. (A) A network in which nodes are labeled according to some
  existing inherent ordering or hierarchy, which identifies an
  inherent directionality.  (B) In any given feedback loop, arrows
  point in the direction of increasing labels, i.e. along the inherent
  directionality, with probability $\gamma$ (blue arrows) or against
  it with probability $1-\gamma$ (red arrows).  (C) Example of
  networks with $\gamma=1/2$ (random directionality) and with
  $\gamma=1$ (perfect directionality).
  }
\label{fig_model}
\end{center}
\end{figure}

In this paper, we present a simple model relating an assumed degree of
inherent directionality with the statistics of feedback loops in 
networks. Our model depends on a single parameter, $\gamma$, defined
as the probability of any link in the network to point along the
inherent direction (see Fig. \ref{fig_model}). An analytical calculation allows us to predict the
fraction $F(k)$ of loops of length $k$ which are feedback loops.  We
show that, as long as there exist a inherent directionality, i.e. as
long as $\gamma \neq 1/2$, the fraction of feedback loops $F(k)$ of
any loop lengths $k$ --for which we provide analytical estimations--
is much smaller than it would be in network randomizations.

To test the model predictions against empirical data, we scrutinize a
number of empirical biological, ecological, and also socio-technological
directed networks.  For each of these empirical networks, we perform an
extensive computational study of the number of structural and feedback
loops it includes.  In nearly all the networks we analyzed, we find
that $F(k)$ is dramatically smaller than in randomizations of the same
networks.  Remarkably, the model reproduces the curves $F(k)$ with
good precision for all the empirical networks we studied, just by fitting
its only free parameter, quantifying the degree of inherent
directionality.

Furthermore, we introduce a method to directly estimate the degree of
directionality in any given network by employing topological
information only. The resulting measurement for each specific network
correlates quite well with the directionality parameter employed to
obtain the fit for the statistics of feedback loops. We also verify
that our results are robust against network subsampling or lack of
knowledge of existing connections.  Therefore, we conclude that the
lack of feedback loops stems from the existence of a inherent
directionality in empirical networks.

\section*{Results}
\subsection*{Counting loops in empirical networks}  

We analyzed a large set of empirical biological, ecological and
socio-technological directed networks taken from the literature (for
the complete list see Supplementary Information S1). We excluded from
our analyses un-directed networks and tree-like networks with no
single loop of any size. Self-loops --being unrelated to inherent
directionality-- have not been taken into account.  For each network
and each loop-length $k$, we exhaustively counted the number of
structural loops and the fraction of them which are also feedback
loops, ${{F}}(k)$.  We remark that knowledge
of the  hierarchical level of each node (if any) is not necessary for this computation.

  From a computational perspective, counting loops is a non-polynomial
  (NP) hard problem, thus becoming an unfeasible task for large
  network sizes. For this reason, previous studies often used less
  computationally-expensive proxies –such as the Estrada index
  \cite{Estrada-index} or analytical estimations for large network
  sizes \cite{Ginestra-loops}– to estimate the amount of loops in
  empirical networks.  Despite the non-polynomial nature of the
  problem, present computer power allows us to count loops up to
  reasonably-large sizes by using an efficient breadth first algorithm
  (see Supplementary S2 for more information on the algorithm).

We compared the measured fraction of feedback loops $F(k)$ with two
different randomizations of the same network. The first one --that we
term directionality randomization (DR)-- preserves the existing links,
but fully randomizes their directions. The second one -- configuration
randomization (CR)-- randomizes both links and directions, but
preserving the in and out connectivity of each single node
\cite{configuration-model} (see Supplementary S3).

Our results, shown in Fig. \ref{fig_fit_loops_ecobio}, exhibit a clear
trend: the fraction of feedback loops of any length $k$ is much
smaller in biological and ecological networks than would be expected
for any of the two different randomizations.  Let us caution that
randomly wired networks of finite size can exhibit small statistical
deviations from the large-size limit $\gamma=1/2$.

The total number of feedback loops --not just its fraction-- is also
severely reduced with respect to network randomizations in all the
considered biological and ecological networks, as firstly noted in
\cite{bianconi2008} (see Supplementary Fig. S2).  These trends are not
so evident for socio-technological networks; while all of the
considered networks have a smaller fraction of feedback loops than
their directionality randomizations, some of the social ones
(e.g. ``twitter followings'' and ``political blogosphere'') have a
larger $F(k)$ than configurational randomizations.

We now test the predictions of our probabilistic model against the
empirically measured values of $F(k)$ in all empirical networks.  For each
of the analyzed empirical networks we consider loop lengths ranging
from $k=3$ to maximum values up to $k=12$, determined by computational
capabilities and depending crucially on network size and
connectivity. For each network, we estimate the value of the
directionality parameter $\gamma$ which best describes the observed
fraction of feedback loops via an unweighted least-square fit of $\log
F(\gamma,k)$ as a function of $k$.

Results are summarized in Fig.\ref{fig_fit_loops_ecobio}.  The model
reproduces remarkably well empirical data for all loop lengths by
fitting the only free parameter.  In some cases, such as for the
neural connectivity (C. elegans) network, the agreement between
empirical data and model predictions is quite impressive, while
significant deviations are observed in some other cases for small loop
lengths, $k \le 4$.  In particular, the worst agreement is obtained
for the Coachella valley foodweb. However, this network, with only
$29$ nodes, is the smallest in the dataset, so that it can deviate
significantly from statistical predictions and it has been previously
reported to be anomalous from other viewpoints \cite{anomalous}.  In
some cases, such as the N.E. Shelf foodweb and the two considered
transcription regulatory networks ({\it E. coli} TRN and Yeast TRN),
$\gamma > 0.999$ indicating a rather extreme level of inherent
directionality (see Table \ref{table_corr}). We obtained similar
results for other empirical networks with very few loops (listed in
Table \ref{table_corr} as well), providing additional support to our conclusion.

\begin{figure}[ht!bp]
\begin{center}
\includegraphics[width=1\textwidth]{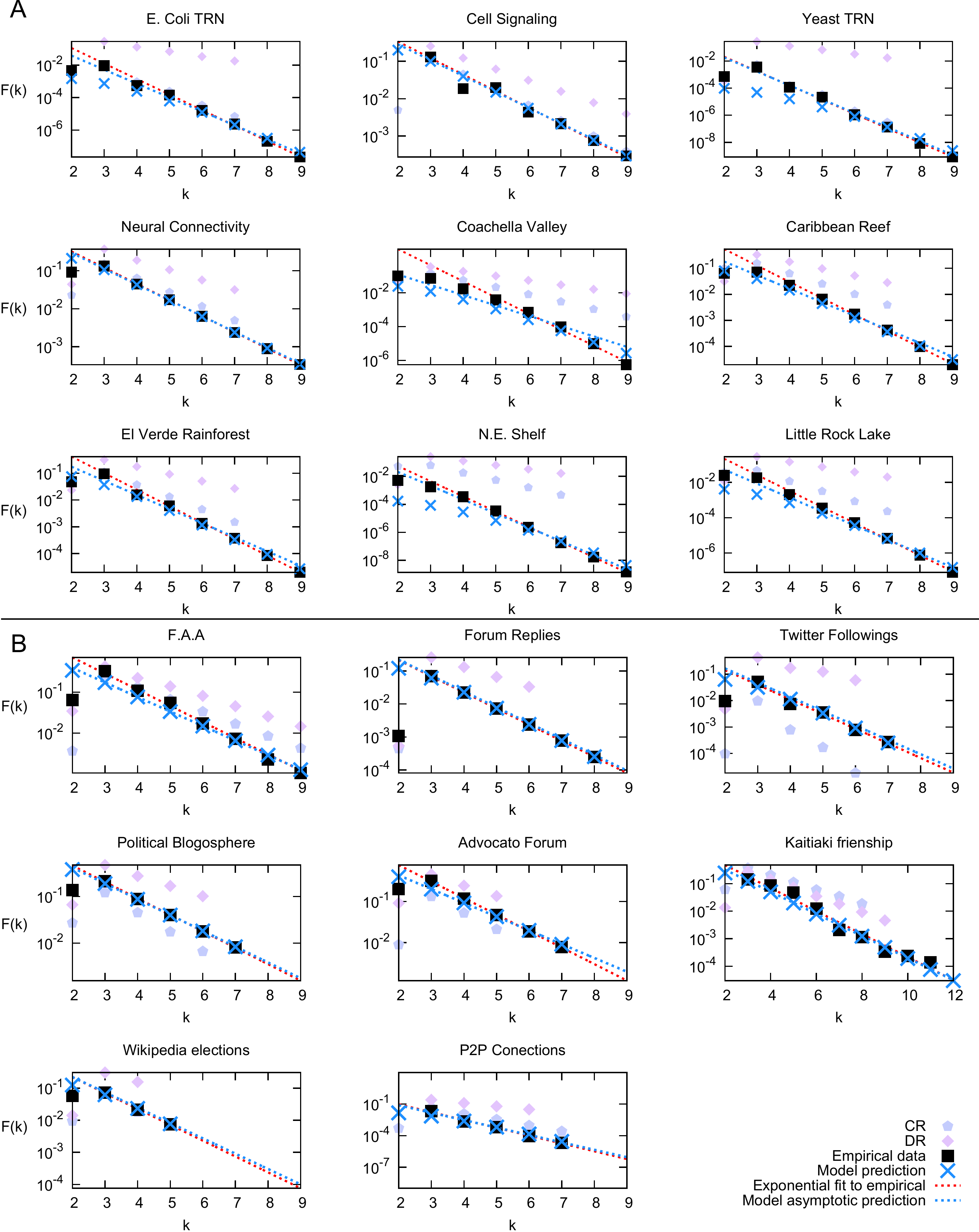}
\caption{
{\bf Fraction of feedback loops, $F(k)$, as a function of the
    loop length, $k$, in empirical networks}.  Black squares
  correspond to empirical data and red dashed lines stand for fits of
  the empirical data to an asymptotic exponential curve (fit done
  using data for $k>4$).  Pale blue pentagons stand for
  configurational randomizations and pale pink diamonds for
  directionality randomizations. Blue crosses mark the best fit of our
  probabilistic model (the parameter $\gamma$ has been fitted using a
  least-squares method to $\log F(k)$ versus $k$).  The resulting
  optimal $\gamma$ values for the different networks are compiled in
  Table \ref{table_corr}.  Blue dashed lines correspond to the
  asymptotic analytical estimate of Eq.(\ref{main}) for the
  corresponding $\gamma$.  Notice the closeness between the
  exponential fit to empirical data and the analytical prediction.
  }
\label{fig_fit_loops_ecobio}
\end{center}
\end{figure}

As the model predicts an asymptotic exponential decay of $F(k)$ as the
loop-length $k$ increases, we have performed --for each particular
network-- a fit of the empirical data (for $k>4$) to an exponential
function (see dashed red lines in Fig. \ref{fig_fit_loops_ecobio}).  In
this case, the quality of the fit of $\log F(k)$ versus $k$ can be
assessed via a linear regression coefficient, $r$.  Obtained values of
$r^2$ (Table \ref{table_corr}) are larger than $0.99$ in all cases except
one --the Mammalian cell signaling network, for which $r^2=0.973$--
indicating that even for relatively small loop-lengths the predicted
asymptotic exponential decay holds.  Furthermore, each of these
exponential fits is very close to its corresponding
analytically-obtained asymptotic result, Eq. (\ref{main}) (blue
discontinuous lines in Fig. \ref{fig_fit_loops_ecobio}). 
In the few cases in which the analytical asymptotic prediction breaks
down (see Methods) the blue dashed lines correspond to a fit of the model 
data for $k\le 4$. This shows that the asymptotic expression is reasonably 
accurate even for rather short loops.

We conclude this section with a remark on the possible impact of
unknown links. Our knowledge of biological and technological networks
is often incomplete and it is important to assess how this fact may
affects our analyses. To test the robustness of our framework, we
mimicked the effect of undersampling of empirical networks by
eliminating a fraction of the links at random, and repeated the
analysis above. While this operation clearly affects the number of
links, the conclusions of our model (in particular the fitted value of
$\gamma$) are very weakly modified even when a relatively large
fraction of nodes is removed $20\%\sim 50\%$. Details are presented in
Supplementary Information S5 and Supplementary Fig. S3.

\subsection*{Measuring the degree of directionality of empirical networks}

The directionality parameter $\gamma$ in the probabilistic model
represents how strongly the hypothesized hierarchical ordering affects
the direction of the links in the network; $\gamma=1$ (and also
$\gamma=0$) reflect perfect directionality while $\gamma=1/2$
corresponds to random directionality. In the previous section,
$\gamma$ has been inferred from the statistics of feedback loops.

We now propose an algorithm to directly measure the degree of
directionality of a network from its topology. Similar methods have
been proposed for this purpose
\cite{lagomarsino,Yu_Gerstein,computation_GRN}.  All of them are able
to extract a hierarchical ordering from a given network and classify
nodes into a few discrete levels. Instead, the method we propose
produces more refined orderings, being able to resolve possible
degeneracies between the coarser levels produced by previous methods
(see \cite{AmNat}).

Our method is inspired by algorithms for determining trophic levels in
food webs, but is applicable to any directed network; it can be also
seen as a way to infer a ``hidden variable'' from network topology
\cite{boguna_PRL}.  As customarily done with food webs, one identifies
``basal nodes'' as those having zero in-connectivity, i.e. with no
link pointing to them. In the possible case in which no basal node
exists, we progressively identify sets made out of two, three... nodes
which --taken as a unique coarse-grained node-- are basal, i.e. no
external node points to any node in the set.

Basal nodes obtained in this way are placed at the lowest level of the
hierarchical ordering, $l=0$.  Then, the level of the remaining nodes
is defined as the average of the trophic level of all nodes pointing
to it (its preys in food webs) plus $1$:
\begin{equation}
l_j = 1+ \frac{1}{k_j} \sum_i A_{ij} l_i ,
\label{iteration}
\end{equation}
where $k_j$ is the in-connectivity of node $j$, $A_{ij}$ is the
connectivity or adjacency matrix and $l_j$ is the hierarchical level
of node $j$. The conditions (\ref{iteration}) define a set of linear
equations in the unknown $l_j$'s that can be solved using standard
algebraic methods. Notice that, while with existing methods
\cite{lagomarsino,Yu_Gerstein,computation_GRN} hierarchical levels
associated to nodes are integer numbers, here they are in general real
numbers. Further details, examples and applications of this method
will be published elsewhere.

Using the hierarchical ordering resulting from applying the algorithm
above, it is straightforward to compute the fraction of links pointing
from lower to higher hierarchical levels, i.e. aligned with the
inherent directionality. We call this fraction ``current parameter'',
$\chi$. In the limit of perfect
feedforwardness one expects $\chi=1$, while
in the absence of a well-defined directionality $\chi \approx 1/2$
(apart from small deviations due to finite-size effects).

\begin{figure}[t]
\begin{center}
\includegraphics[width=9cm]{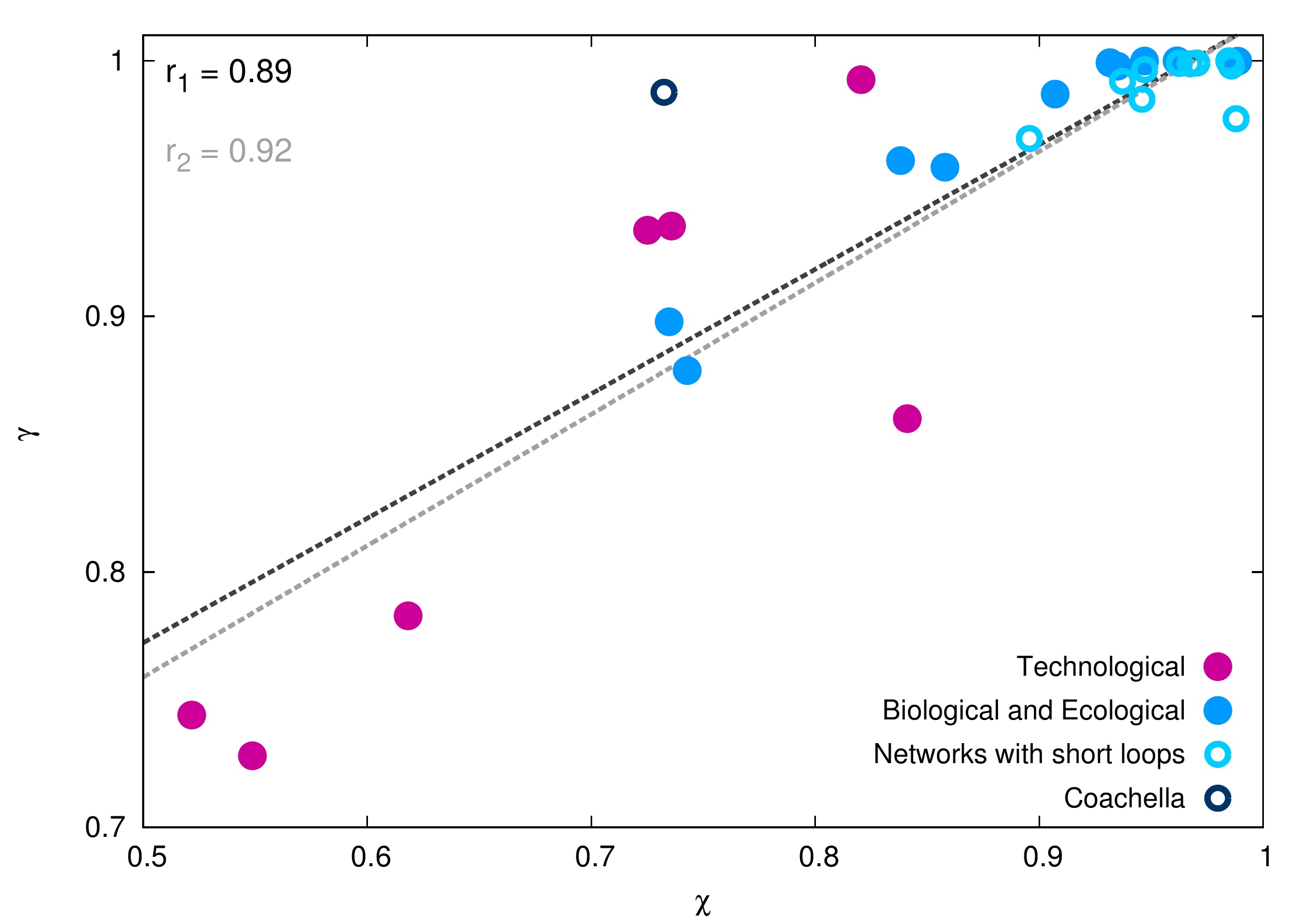}
\caption{
{\bf Correlation between inferred and explicitly measured
    levels of directionality}.  Scatter plot of the optimal values of
  the directionality parameter $\gamma$ plotted against the current
  parameter $\chi$.  Values of either $\gamma$ or $\chi$ close to $1$
  reflect a high degree of directionality while smaller values close
  to $1/2$ imply that link directions are nearly uncorrelated with
  directionality. The value of the linear correlation coefficient is
  $r=0.89$ or $r=0.92$ depending on whether the outlier ``Coachella
  Valley'' small network is included or not. The corresponding best
  fits are $\gamma= 0.487  \chi + 0.529$ and $\gamma =0.514 \chi +
  0.502$, respectively.
  }
 \label{fig_correlation}
\end{center}
\end{figure}

Our results are summarized in Fig. \ref{fig_correlation}. They
clearly show that all the considered biological, ecological, and also
--to much lesser extent-- socio-technological networks exhibit some
degree of hierarchy, $\chi > 1/2$. More remarkably, the explicitly
measured values of $\chi$ correlate quite well with the fitted value
of the directionality parameter $\gamma$ in the set of networks under
study.  This correlation implies that the free parameter we use to fit
the directional model is consistent with a direct measure of
directionality (current) in the same networks.

\section*{Discussion}
While the crucial role of feedback loops in determining dynamical
properties of complex networks has been widely recognized in the
literature, their statistics remained scarcely studied.  Some
exceptions are Refs. \cite{thieffry98,yeast_structure} where,
respectively, the under-representation of long feedback loops in the
{\it E. coli} gene regulatory network and the over-representation of
short feedback loops in the {\it S. cerevisae}'s one were first
noticed, as well as \cite{bianconi2008} where the statistics of the
total number of feedback loops in complex networks was studied.
 
We have tackled the problem of exhaustively counting the number of
structural loops and feedback loops in a variety of biological,
ecological, and socio-technological networks. We then compared these
numbers with those in randomized versions of the same graphs, where
other basic structural features (such as total number of nodes, number
of links, connectivity of each link, etc.) were preserved. In all the
analyzed biological and ecological networks we find a dramatic
reduction of the fraction of loops which are also feedback loops with
respect to random expectations. This effect is much milder in
socio-technological networks.

We hypothesize that the (empirically observed) lack of feedback loops
stems from the existence of an inherent directionality. To investigate
this conjecture, we have constructed a simple computational model in
which an inherent network directionality --quantified by a
directionality parameter $ \gamma$-- is built in.  For this model we
are able to analytically compute the fraction of feedback loops of any
given length as a function of $\gamma$. Our main result is that this
intrinsically directional model can reproduce quite well empirical
curves of the fraction of feedback loops of any length by just tuning
its only parameter $\gamma$. For example, for some networks such as
the neural connectivity network, empirical results fall in a
nearly-perfect way on top of the model curve for all loop-lengths with
amazing accuracy.  The quality of the results is even more remarkable
if we consider that our model assumes a number of simplifications that
are by no means trivial. For instance, the model neglects any
correlation or relation among different loops: each loop is treated
separately, while in empirical networks, especially if they have broad
connectivity distribution functions, typically loops are not
independent as they can share some nodes. In particular, hubs are
statistically more likely than other nodes to take part in
loops. Furthermore, node degree and position in the network hierarchy
could well be correlated in empirical networks, while such an hypothetical
correlation is just neglected by our simple model. These effects could
be responsible for the small departures of empirical data from our
model predictions.

It is even more remarkable that the optimal value of the
directionality parameter $\gamma$ --derived from the statistics of
loops-- correlates quite well with the current parameter, $\chi$,
computed by quantifying the network ``stratified'' architecture or
degree of directionality. These two measures of network inherent
directionality are quantitatively different but they are strongly
correlated.

It is interesting to recall that the first model of food
  web architectures \cite{Cohen_1} did include a perfect
  directionality and thus complete absence of feedback loops, while
  more recent models (see
  e.g. \cite{Williams_niche_model,allesina_model,Staniczenko_selecting})
  allow for some small degree of backward edges, enabling directed
  loops to appear.

Our finding is similar in spirit to the remarkable observation by
Mayaa'n {\it et al.}  that biological networks display a kind of
antiferromagnetic ordering --meaning that contiguous links have a
statistical tendency to point in opposite directions-- causing a
depletion of feedback loops which they claim lead to an enhancement of
network stability \cite{Cecchi-PNAS}. Instead, our hypothesis here is
that the absence of feedback loops is a byproduct of a more inherent
feature of networks: the existence of a preferred
directionality. Indeed, by employing a method inspired on how trophic
levels are identified in food webs, we have been able to identify
--just by looking at the network structure-- an objectively measured
correlate of the fitted directionality parameter.  Similarly, in a
recent work, it is claimed that long loops are over-represented in
biological networks \cite{Louzoun}. The origin of the apparent
conflict with our results can be tracked down to the different
definition of loops employed in \cite{Louzoun}, where only ``minimal
loops'' (see \cite{Louzoun} for a definition) are considered rather
than the exhaustive enumeration of all loops we perform here.

Summarizing, our results show that the existence of an inherent
directionality constitutes a simple yet satisfactory parsimonious
explanation for the empirically observed lack of feedback loops in
biological and ecological networks.

\section*{Methods}
\subsection*{Network Directionality model}
Let us consider a network consisting of $N$ nodes and $L$ directed
links and imagine that the fraction of loops which are also feedback
loops, $F(k)$, is known.  We now aim at constructing a probabilistic
model able to predict the empirically-measured function
$F(k)$. The model consists in taking the empirical network under
consideration and randomizing the direction of each single link with
the constraint that some degree of inherent directionality exists. We
therefore assume that nodes can be characterized by an index or
coordinate $i=1\dots N$ representing their position along the
directionality axis. As a convention, we choose higher nodes in the
hierarchy to have larger labels, as shown in Fig.
\ref{fig_model}A. A direction to each existing link is (re-)assigned
as follows (see Fig. \ref{fig_model}B): a link is set to point from
a lower label to the higher one, with probability $\gamma$, where the
``directionality parameter'' $\gamma$ satisfies $0\le\gamma\le1$. With
the complementary probability $1-\gamma$ the link points against the
inherent directionality.  In particular, $\gamma=1$ (or $\gamma=0$)
stands for perfect inherent directionality, while for $\gamma=1/2$,
the inherent directionality does not affect the direction of the
links.

Our goal is to analytically estimate the expected value of $F(k)$ for
any given loop length $k$ as a function of the only parameter. To make
progress, we consider loops independently, i.e. we neglect possible
correlations between for example loops having common links in a same
network.  We also neglect the impact of possible heterogeneities in
the distribution of loops across hierarchical levels.  In the case of
empirical networks we are interested in, we shall assume these as working
hypotheses, whose validity will be tested {\em a posteriori} by
comparing our results against data.

Under these assumptions, we focus on a specific loop of arbitrary length $k$ (see
Fig. \ref{fig_model}). Without loss of generality, we re-label the node indexes
onto the integer numbers $1\dots k$ by preserving the ordering,
i.e. we label the node having the lowest index in the loop with $1$,
the second lowest with $2$ and so on.  In this way, the loop is
associated with a permutation $\{n\}=n_1,n_2\dots n_k$, where $n_i$ is
the label of the $i-th$ node in the loop. Formally, we define
$n_{k+1}=n_{1}$ to ensure that the loop is closed.

Under the assumptions above, we consider that all the $k!$ possible
loop permutations are equally likely to be found. In this way, the
maximum number of feedback loops is expected to occur for
$\gamma=1/2$, for which the two directions are equi-probable. In this
case, $F(k)= 2^{1-k}$ as only $2$ out of the possible $2^k$ loops of
length $k$ are feedback loops.  In a more general case, the
probability of a given loop to be a feedback loop depends on the
distribution of {\em the number of ascents}, i.e. the number $A(l,k)$
counting how many permutations of the basic sequence of length $k$ are
such that $n_i<n_{i+1}$ holds for exactly $l$ distinct values of
$i$. For a non-periodic sequence, i.e. without establishing any
relation between $n_k$ with $n_1$, the solution to this problem is
given by the so-called Eulerian numbers (see e.g. \cite{comtet}
chapter 6 or \cite{Eulerian}).  Since loops are closed, we need to
generalize the concept of Eulerian numbers to the periodic or cyclic
case, i.e. we need to count the number of ascents in a generic closed
loop, which we call ``cyclic Eulerian numbers'', $A(l,k)$. 
Further in this section we prove a recursion relation
\begin{equation}\label{recur}
(k-1) A(l,k) = k[(k-l) A(l-1,k-1)  + l A(l,k-1) ]
\end{equation}
which generalizes a similar relation for standard Eulerian numbers
(see e.g. \cite{comtet}) and which allows us to recursively find all
cyclic Eulerian numbers. Notice, in particular, that
$A(0,k)=A(k,k)=0\ \forall k$ as it is clearly impossible to have all
ascents/descent in a closed loop.  Examples of cyclic Eulerian numbers
for values of $k$ up to $9$ are also presented later in Methods.

\begin{figure}[ht!bp]
\begin{center}
\includegraphics[width=9cm]{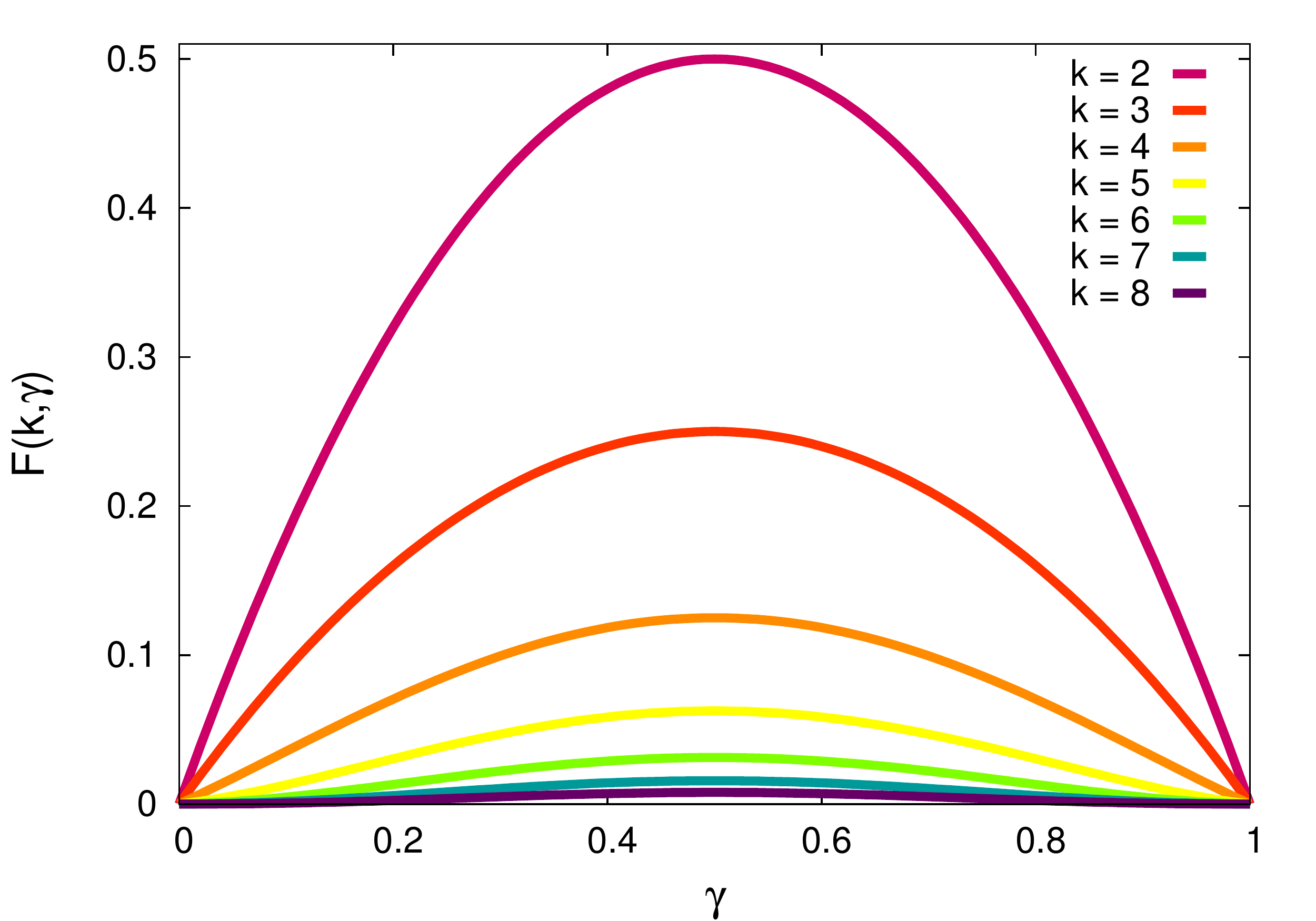}
\caption{
{\bf Fraction of feedback loops, $F(k)$, versus the
    directionality parameter $\gamma$}. $F(k)$ has a maximum at
  $\gamma=1/2$, for which all link directions are randomly set, giving
  rise to the largest possible fraction of directed loops.  On the
  other side, $F(k)$ vanishes for $\gamma=0$ and for $\gamma=1$ as
  expected. Notice also that the curves are symmetric around
  $\gamma=1/2$ and that for values of $\gamma$ different from $1/2$
  one has a directionality-induced lack of feedback loops.
  }
  \label{fig_gamma}
  \end{center}
\end{figure}

The expected fraction $F(k,\gamma)$ of loops of length $k$ which are
feedback loops can be expressed as
\begin{equation}\label{phi}
  F(k,\gamma)=\sum_{l=0}^k \frac{A(l,k)}{k!} 
  \left[\gamma^l(1-\gamma)^{k-l}+\gamma^{k-l}(1-\gamma)^{l}\right],
\end{equation}
where the two terms in square brackets account for the two different
possible orientations of a feedback loop. The function $F(k,\gamma)$
is plotted in Fig. (\ref{fig_gamma}) as a function of $\gamma$ for
different values of $k$. $F(k,\gamma)$ is symmetric by exchanging
$\gamma$ by $1-\gamma$, corresponding to reversing the direction of
the inherent directionality. Note that imposing the
  normalization condition $\sum_l A(l,k)=k!$, one can easily retrieve
  from Eq.(\ref{phi}) the probability $F(k,1/2)=2^{1-k}$ in the
  limiting case $\gamma=1/2$.
  
The exact expression of Eq. (\ref{phi}) can be
approximated in the asymptotic limit of large $k$ and $\gamma$ not too
small (see Supplementary S6) by the expression
\begin{equation}\label{main}
{F}(k,\gamma)\approx 2 \exp\left\{
  \frac{k}{2}\log[\gamma(1-\gamma)]+\frac{k}{24}
  \log^2\left(\frac{\gamma}{1-\gamma}\right)
\right\}.
\end{equation}
Eq.(\ref{main}) predicts that the fraction of feedback loops decays
exponentially with the loop length $k$ with an amplitude factor $2$
and with an exponential constant which depends on $\gamma$.

\subsection*{Number of ascents and cyclic Eulerian numbers}\label{sec:eulerian_numbers}
Let us consider a loop of length $k$, formed by a closed chain of $k$
nodes and $k$ edges, and let us label the nodes with numbers from $1$
to $k$.  We consider all the $k!$ possible permutations of labels and
aim at computing the number $A(l,k)$ of such permutations including
$l$ ascents, i.e. permutations in which exactly $l$ labels in the
sequence are immediately followed by a larger one. The first goal is
to verify that the $A(l,k)$'s satisfy a simple recurrence relation,
similar to that obeyed by standard Eulerian numbers (see
e.g. \cite{comtet} chapter 6 and \cite{Eulerian}).  To establish such
a relation, let us first observe that the number of ascents does not
depend on the specific ordering/permutation within a cycle. For
instance the permutations $123(1)$, $231(2)$ and $312(3)$, which
correspond to three different ways of labeling the cycle $A
\rightarrow B\rightarrow C \rightarrow A $, have the same number of
ascents ($2$, in this example).  Therefore $A(l,k) = k C(l,k)$ where
$C(l,k)$ corresponds to the number of ascents in the case in which the
symmetry has been broken and one specific label has been chosen to be
at the opening and closing extremes of the representation above. Now
we look for a recurrence relation for $C(l,k)$, for which we need to
express $C(l,k)$ as a function of $C(j,k-1)$, where $j= l$ or
$j=l-1$. These correspond to two different cases that can occur when a
new node is inserted in a loop to create a one-step larger
sequence. If the node is inserted where there was an ascent, it simply
replaces the previous one, so that the number of ascents remains
unaltered. If it is inserted where there was a descent, a new ascent
is created, so that $l$ is increased by one. These two possibilities
can be summarized in the recursive equation
\begin{equation} \label{eq_c}
C(l,k)= C(l,k-1)  l + C(l-1,k-1) (k-1-(l-1)),
\end{equation}
where the two cases above have been weighted with the number of
ascents and descents, respectively. Eq. (\ref{recur}) follows
straightforwardly from Eq. (\ref{eq_c}) and $A(l,k) = k
C(l,k)$. Specific values for $k \leq 9$ obtained by iterating the
recursive formula are shown in Table I.

\bibliographystyle{naturemag}
\bibliography{loops}

\section*{Acknowledgments:} 
We are grateful to S. Johnson for a long term pleasant collaboration
and to Y. Moreno for providing us with network data. We thank
A. Bernacchia for suggesting the analogy with Eulerian numbers and
P. Moretti and J. Hidalgo for a critical reading of the manuscript.

\section*{Author Contributions:}
Conceived and designed the work: VDG SP MAM.  Performed the
experiments: VDG SP MAM.  Analyzed the data: VDG SP MAM.  Contributed
reagents/materials/analysis tools: VDG SP MAM.  Wrote the paper: VDG
SP MAM.

\section*{Additional Information}
The authors declare no competing financial interests.

\bibliographystyle{naturemag}


\clearpage

\section*{TABLES}

\begin{longtable}[h!tp]{| @{}l|   c |c |c|}
\hline
Network & $r^2$ & $\gamma$ & $\chi$\\
 \hline
\hline
{\it E. coli} TRN & 0.995 & 0.999 & 0.9316 \\
Cell Signaling & 0.973 & 0.888 & 0.7348 \\
Yeast TRN & 0.997 & 1.000 & 0.9887 \\
Neural Connectivity & 1.000 & 0.879 & 0.7429 \\
Coachella Valley & 0.984 & 0.988 & 0.7325 \\
Caribbean Reef & 0.997 & 0.958 & 0.8579 \\
El Verde Rainforest & 0.997 & 0.961 & 0.8381 \\
N.E. Shelf & 0.997 & 1.000 & 0.9470 \\
Little Rock Lake & 0.999 & 0.998 & 0.9350 \\
Lough Hyne & 0.980 & 0.999 & 0.9616 \\
Weddell Sea & 0.992 & 0.987 & 0.9072 \\
\hline
F.A.A. & 0.998 & 0.783 & 0.6183 \\
Forum Replies & 1.000 & 0.935 & 0.7359 \\
Twitter Followings & 0.984 & 0.967 & - \\
Political Blogosphere & 1.000 & 0.744 & 0.5217 \\
Advocato Forum & 0.999 & 0.714 & 0.5487 \\
Kaitiaki friendship & 0.975 & 0.860 & 0.8411 \\
Wikipedia elections & 0.998 & 0.934 & 0.7252 \\
P2P Connections & 0.991 & 0.993 & 0.8205 \\
\hline
\hline
Everglades & 0.979 & 0.999 & 0.9673 \\
Mangrove Estuary & 0.998 & 0.999 & 0.9704 \\
Mondego Estuary & 1.000 & 0.992 & 0.9373 \\
Skipwith & 0.000 & 0.997 & 0.9471 \\
Human TRN & 0.984 & 0.999 & 0.9626 \\
Mouse TRN & 1.000 & 0.970 & 0.8957 \\
Ownership & 1.000 & 0.977 & 0.9880 \\
Tuberculosis TRN & 1.000 & 0.998 & 0.9858 \\
{\it B. subtilis} TRN & 1.000 & 0.985 & 0.9459 \\
\hline
\caption{
{\bf Quantification of network directionality.} First and
  second columns: values of the linear correlation coefficient $r^2$
  and of the fitted parameter $\gamma$, respectively, for
  the linear fit of $\log {{F}}(k)$ versus $k$ with
  Eq. (\ref{phi}) for the considered networks. Third column: measures of
  the current parameter $\chi$ from the network structure (large
  values of $\chi$ indicate high levels of hierarchy and thus of directionality).  
  Networks below the central double line are those with only a small
  number of short loops, i.e. not having any loop larger than $k=6$. In
  the case of the Skipwith network, the value of $r^2$ is absent as we
  could not compute long enough loops to observe the exponential
  decay. In the Twitter followings network the value of $\chi$ could not
  be computed due to computational limitations. 
}
\label{table_corr}
\end{longtable}

\clearpage

\begin{longtable}[h]{|l|c|c|c|c|c|c|c|c|c|}
 \hline
k \ l & 0 & 1 & 2 & 3 & 4 & 5 & 6 & 7 & 8 \\
\hline
\hline
1 & 1 &  &  &  &  &  &  &  &    \\ 
\hline
2 &  & 2 &  &  &  &  &  &  &    \\ \hline
3 &  & 3 & 3 &  &  &  &  &    &  \\ \hline
4 &  & 4 & 16 & 4 &  &  &    &  &  \\ \hline
5 &  & 5 & 55 & 55 & 5 &  &  &  &   \\ \hline
6 &  & 6 & 156 & 396 & 156 & 6 &    &  &  \\ \hline
7 &  & 7 & 399 & 2114 & 2114 & 399 & 7 &  & \\ \hline
8 &  & 8 & 960 & 9528 & 19328 & 9528 & 960 & 8 &    \\ \hline
9 &  & 9 & 2223 & 38637 & 140571 & 140571 & 38637 & 2223 & 9  \\
 \hline
\caption{Cyclic Eulerian numbers $A(l,k)$,where $k$ is the size of the loop and $l$ the number of ascents.}
\end{longtable}

\end{document}